\documentclass[pra, 12 pt]{revtex4}
\usepackage[english]{babel}
\usepackage{amsmath}
\usepackage{epsfig}

\begin{document}

%\draft
\title{\bf IDENTITY OF ELECTRONS AND IONIZATION EQUILIBRIUM}
\author{V.B. Bobrov $^1$, S.A. Trigger $^{1,2}$,  W. Ebeling $^2$}
\address{$^1$ Joint\, Institute\, for\, High\, Temperatures, Russian\, Academy\,
of\, Sciences, 13/19, Izhorskaia Str., Moscow\, 125412, Russia;\\
email:\, satron@mail.ru;\\
$^2$ Institut f\"ur Physik, Humboldt-Universit\"at zu Berlin,
Newtonstra{\ss}e 15, D-12489 Berlin, Germany}

\begin{abstract}
It is perhaps appropriate that, in a year marking the 90th
anniversary of Meghnad Saha seminal paper (1920), new developments
should call fresh attention to the problem of ionization
equilibrium in gases. Ionization equilibrium is considered in the
simplest "physical" model for an electronic subsystem of matter in
a rarefied state, consisting of one localized electronic state in
each nucleus and delocalized electronic states considered as free
ones. It is shown that, despite the qualitative agreement, there
is a significant quantitative difference from the results of
applying the Saha formula to the degree of ionization. This is
caused by the fact that the Saha formula corresponds to the
"chemical" model of matter.\\

PACS number(s): 52.25.Jm, 52.25.Ya, 05.30.Fk, 34.80.Dp

\end{abstract}

\maketitle

The Saha equation [1] derived in 1920 plays a fundamental role in
all fields of plasma physics [2], and defines the ratio of
electrons, ions, and atoms at given thermodynamic parameters in an
ideal (or weakly nonideal) mixtures of these particles. The Saha
equation also plays an important role in cosmology, defining the
recombination transition in models of the early Universe [3,4].

The Saha equation was derived based on the matter model in which
bound states of particles are considered as new composite
particles with a priory specified structure and properties. This
means, that real matter which should be considered in most cases
as a quasi-neutral non-relativistic many-body system consisting of
interacting nuclei and electrons ("physical" model of matter [5])
can be considered under certain conditions as a system of atoms or
molecules. Furthermore, atoms or molecules in such a model, being
in essence bound states of a certain finite number of electrons
and nuclei, represent a form of quasiparticles [5]. A statistical
description under such consideration requires a strong assumption
which is a basis of the so-called "chemical" model of matter,
i.e., fundamental (electrons and nuclei) and composite particles
should be considered equivalently (see, e.g., [5]). Attempts to
justify this assumption are not terminated to date (see, e.g., [6]
and references therein). In our opinion, this is caused by two
important factors. One situation is associated with that the
"chemical" model has reasonable physical grounds when considering
matter in the rarefied state. However, in describing matter at
high densities, the concept on atoms, molecules, and other
composite particles loses meaning. This means that the statistical
description of the "chemical" model should imply the formalism of
the "appearance" or "disappearance" of composite particles (atoms
and molecules). In this case, the case in point is their principal
presence or absence, rather than their small number which is
assumed by the Saha formula. Another situation is associated with
the consideration of the identity (indistinguishability) of
fundamental particles, first of all, electrons (see, e.g., [7]),
whose consequence is the Pauli principle for electrons [8]. The
point is that the number of chemical potentials in the statistical
description of matter in the "physical" model (which is primary)
is defined by the number of types of distinguishable particles
(see, e.g., [9]). In particular, for the pure matter consisting of
electrons and nuclei of the same type, the theory implies two
chemical potentials, i.e., the chemical potential of electrons and
the chemical potential of nuclei. In this case, three chemical
potentials are put into consideration in the "chemical" model,
i.e., the chemical potential of "free" (delocalized) electrons,
the chemical potential of "ions" (or free nuclei), and the
chemical potential of "atoms" consisting of bound electrons and
nuclei. Thus, if the problem of the introduction of the chemical
potential of "atoms" (see above) is not considered, the
possibility of replacing the initial chemical potential for all
electrons by the chemical potential of only "free" electrons
remains unclear. In our opinion, the only possible justification
of such a step from the viewpoint of indistinguishability of
electrons is the recognition of the fact that, although electrons
themselves are indistinguishable, electronic states are
distinguishable (see, e.g., [10]). In particular, this is valid
for the distinguishability of localized ("atomic", "molecular",
etc.) and delocalized ("free") electronic states. This
circumstance formally offers the possibility of "different"
mathematical descriptions of localized and delocalized electronic
states using various chemical potentials. From this point of view,
the actual reason of the development of the "chemical" model of
matter is caused by difficult construction of the general theory
of systems with Coulomb interaction. The point is that it is
necessary to uniformly describe both localized electronic states
which are characterized by the strong electron--nucleus
interaction and delocalized electronic states which are quite
adequately described within the perturbation theory with respect
to interparticle interaction (see, e.g., [9]). In this situation,
the problem of the quantitative relation between the results of
applying the "chemical" and "physical" models to describe the
ionization equilibrium in the rarefied gas state, where the
"chemical" model has reasonable grounds, becomes central. In the
case of the "physical" model, we shall proceed from the assumption
on the classical description of the nucleus subsystem, which
corresponds to the Saha formula derivation. The electron identity
principle requires that all electrons of the system should be
described by the uniform chemical potential $\mu_e$. When
considering localized electronic states in the low-density limit,
we restrict the analysis to the one-center approximation which is
a necessary condition of the existence of "atoms" as
quasiparticles. For convergence of statistical sums, the
electron-electron interaction can be considered within the
self-consistent Hartree-Fock approximation [11].

Denoting electron energy levels in localized and delocalized
states by $E_n$ and $\epsilon(q)$, respectively, and assuming that
localized states are take place at each nucleus, for the average
number $\langle N_e\rangle^{(GE)}$ of electrons in the system, we
obtain [12]
\begin{eqnarray}
\langle N_e \rangle^{(GE)}=N_c \sum_n f_e(E_n)+ \sum_q f_e
(\epsilon(q)). \label{S1}
\end{eqnarray}
Here $N_c$ is the full number of nuclei in the system. The
function $f_e(E)=[1+\exp(E-\mu_e)/T]^{-1}$ is the Fermi energy
distribution. The first and second terms on the right-hand side of
(1) yield the average number of electrons in localized $\langle
N_e \rangle^{(loc)}$ and delocalized $\langle N_e
\rangle^{(deloc)}$ states. In this case, Eq. (1) together with the
quasineutrality condition $\langle N_e \rangle^{(GE)} =z_c N_c$
makes it possible, for a given spectrum of electronic states, to
determine the chemical potential of electrons as a function of the
nuclei density $N_c/V$, temperature $T$, and nucleus charge $z_c$,
hence, to determine the degree of ionization $\alpha$ equal to
ratio of the number of delocalized electrons to the number of
nuclei in the system under study,
\begin{eqnarray}
\alpha=\frac{\sum_q f_e (\epsilon(q))}{N_c}. \label{S2}
\end{eqnarray}
Bearing in mind the comparison of this approach with the Saha
formula for the ionization equilibrium in its simplest form, we
make a number of simplifying (but generally speaking, unnecessary
for the general approach described above) approximations. Let us
assume that delocalized states in the low-density limit can be
approximately described by plane waves with $\epsilon (q) =h^2
q^2/2m_e$ (strictly speaking, delocalized states orthogonal to
localized electronic states should be searched, e.g., in the form
of the so-called COPW states [13]). Furthermore, when considering
delocalized electronic states, we suppose that $\mu_e <0$,
$\mid\mu_e/T \mid>>1$, replacing the Fermi distribution by the
Boltzmann distribution, whereas localized states are certainly
described by the Fermi distribution. In this case, the number of
delocalized electronic states in the volume $V$ is $2V \exp
(\mu_e/T)$. Equation (1) for determining the chemical potential is
still transcendental, and the problem of its numerical solution is
related to the necessity of determining the energies $E_n$ of
localized excited ($n>1$) electronic states in the Hartree-Fock
approximation [12].

We will now use the known result based on the solution of the Saha
equation. When determining the degree of ionization for hydrogen
and of some other elements, this result makes it possible to
restrict the analysis to the consideration of only the electron
ground level in the atom and states of the continuous spectrum in
wide temperature and density ranges, disregarding excited
localized states [14]. In this case, putting the excited states
into the calculation slightly affects the accuracy of the
determination of the degree of ionization [15].

Let us perform the further consideration without loss of
generality; for the case of hydrogen, $z_c =1$. Under given
conditions, the ground energy level for the localized electronic
state $E_0\equiv -I$ in the Hartree-Fock approximation coincides
with that for the case of complete disregard of the
electron-electron interaction, when $I=m_e e^4 /2 \hbar^2$ (which
corresponds to the Bohr atom) [12]. In this case, it follows from
(1) that the chemical potential in the approximation under
consideration is explicitly defined. In the case of hydrogen,
$z_c=1$, we obtain the equation for $\mu_e$
\begin{eqnarray}
\mu_e= T  \ln  \frac{1}{2} \left \{-[\exp(-(I/T)+\frac{n_c
\Lambda^3}{2}]+\sqrt{[\exp(-I/T)+\frac{n_c\Lambda^3}{2}]^2+2n_c
\Lambda^3 \exp(-I/T)} \right\}; \nonumber\\
\Lambda = \left( \frac{2\pi\hbar^2}{m_eT} \right)^{1/2}
\qquad\;\qquad \qquad\;\qquad \qquad\;\qquad \qquad\label{S3}
\end{eqnarray}
Based on (3), the degree of ionization is calculated as

\begin{eqnarray}
\alpha=\frac{2}{n_c\Lambda^3}\exp(\mu_e/T)= \qquad\;\qquad \qquad\;\qquad \qquad\;\nonumber\\
\exp(-I/T)\frac {-[1+\frac{n_c
\Lambda^3}{2}\exp(I/T)]+\sqrt{[1+\frac{n_c\Lambda^3\exp(I/T)}{2}]^2+2n_c
\Lambda^3 \exp(I/T)}}{n_c\Lambda^3}\label{S4}
\end{eqnarray}
In deriving (2) and (3), we considered the spin degeneracy factor
for energy levels (which was ignored in [12]).

As is known, the degree of ionization calculated in the same
approximation by the Saha formula [1,14], under the assumption of
ideality of atomic, electronic, and ionic components of ionized
gas, is given by

\begin{eqnarray}
\alpha^{(S)}=2\exp(-I/T)\,\frac{-1+\sqrt{1+n_c\Lambda^3
\exp(I/T)}}{n_c\Lambda^3} \label{S5}
\end{eqnarray}
As it is easy to see, both formulas for the degree of ionization
are identical in the low-density limit under the condition that
value $\gamma\equiv n_c\Lambda^3 \exp (I/T)\ll 1$. In this case
and at a fixed temperature, the degree of ionization is close to
full ionization with the accuracy to first order on the value
$\gamma$,

\begin{eqnarray}
\lim_{n\rightarrow 0, T} \alpha\simeq \alpha^{(S)}\simeq 1.
\label{S6}
\end{eqnarray}
The same result corresponds to the transition to high temperatures
at a fixed density of nuclei.

When passing to low temperatures, at a fixed density, the degree
of ionization is close to zero,
\begin{eqnarray} \lim_{T
\rightarrow 0, n} \alpha\simeq \alpha^{(S)}=0. \label{S7}
\end{eqnarray}
Thus, it is clear that the point $n=0, T=0$ is a singular point at
the degree of ionization $\alpha$, similarly to that the point
$q=0, \omega=0$ is a singular point for the permittivity
$\varepsilon (q, \omega)$ which depends on the wave vector $q$ and
frequency $\omega$ (see, e.g., [16,17]). This means that matter
will be in an atomic state at a fixed low density $n$ in the limit
$T\rightarrow0$. In turn, at a fixed temperature $T$ in the limit
$n\rightarrow 0$, matter will be in a completely ionized state.
Although the points $n=0$ and $T=0$ are practically inaccessible,
the non-permutability of limits (see also [15]), written for the
degree of ionization as
\begin{eqnarray}
\lim_{T \rightarrow 0, n_c\rightarrow 0} \alpha(n_c, T)\neq
\lim_{n_c \rightarrow 0, T\rightarrow 0} \alpha (n_c, T).
\label{S8}
\end{eqnarray}
has important physically observable consequences.

Figures 1 and 2 show the degrees of ionization and chemical
potentials in the physical model (solid lines) and according to
the Saha formula (dashed lines) for identical approximations (the
consideration of only one, i.e., the lowest, localized state and
the consideration of delocalized electrons as free particles).

We note that since ions are originally absent in the physical
model, their average charge, can be additionally defined either as
the number of free electrons per one nucleus not occupied by
electrons,

\begin{eqnarray}
z_i=\frac{\langle N_e \rangle^{(deloc)}}{N_c-\langle N_e
\rangle^{(loc)}/z_c}, \label{S9}
\end{eqnarray}
or as the number of free electrons relative to the total number of
nuclei,

\begin{eqnarray}
z^{\ast}_i=\frac{\langle N_e \rangle^{(deloc)}}{N_c}, \label{S10}
\end{eqnarray}
For hydrogen $z_i=1$, while $z^{\ast}_i =\alpha$ and varies from
$0$ to $1$, being a new statistical variable in the
electron-nucleus system.

As follows from the above calculations, despite the qualitative
agreement of the results of applying the "chemical" and "physical"
models to the degree of ionization in rarefied matter, there is a
rather serious quantitative difference between them even in the
considered simplest case. The advantage of the physical model
consists in enumeration of identity of all electrons in the
system. In fact, in the physical model, electrons are distributed
on all existing electron states, according to the Pauli principe,
in contrast with the chemical model, where electrons at first are
separated on atomic and delocalized states, which are described by
a different way. It is easy to establish a formal similarity
between application of the physical model to plasma ionization in
the present paper and consideration of electrons in semiconductors
with donor and acceptor impurities (see, e.g., [18]).

The approach firstly used in the present paper can lead to very
significant quantitative differences in the calculation of
thermodynamic, kinetic, and electromagnetic properties, since the
degree of ionization (or the number of delocalized ("free")
electrons) is of fundamental importance. Thus, the consistent
development of the "physical" (Coulomb) model is of paramount
problem for studying real matter even in the rarefied state.

\end{document}